\documentclass{emulateapj}
\usepackage{graphicx}
\usepackage{natbib}
\slugcomment{Accepted for ApJ, draft from \today}

\shorttitle{Galactic structure based on ATLASGAL}
\shortauthors{Beuther et al.}

\begin{document}


\title{Galactic structure based on the ATLASGAL 870\,$\mu$m survey}


\author{H.~Beuther\altaffilmark{1}, J.~Tackenberg\altaffilmark{1}, H.~Linz\altaffilmark{1}, Th. Henning\altaffilmark{1}, F.~Schuller\altaffilmark{3}, F.~Wyrowski\altaffilmark{2}, P.~Schilke\altaffilmark{3}, K.~Menten\altaffilmark{2}, T.P.~Robitaille\altaffilmark{4}, C.M.~Walmsley\altaffilmark{5,6},, L.~Bronfman\altaffilmark{7}, F.~Motte\altaffilmark{8}, Q.~Nguyen-Luong\altaffilmark{8}, S.~Bontemps\altaffilmark{9}}
%\affil{Astronomy Department, University of California,
%    Berkeley, CA 94720}

%\author{C. D. Biemesderfer\altaffilmark{4,5}}
%\affil{National Optical Astronomy Observatories, Tucson, AZ 85719}
%\email{aastex-help@aas.org}

%\and

%\author{R. J. Hanisch\altaffilmark{5}}
%\affil{Space Telescope Science Institute, Baltimore, MD 21218}


\altaffiltext{1}{Max-Planck-Institute for Astronomy, K\"onigstuhl 17,
              69117 Heidelberg, Germany, beuther@mpia.de}
\altaffiltext{2}{Max-Planck-Institute for Radiostronomy, 
              Auf dem Hügel 71, 53121 Bonn, Germany}
\altaffiltext{3}{University of Cologne, Z\"ulpicher Str.~77, 50937 K\"oln, Germany}
\altaffiltext{4}{Harvard-Smithsonian Center for Astrophysics, 60 Garden Street, Cambridge, USA}
\altaffiltext{5}{Osservatori Astrofisico di Arcetri, Largo E.~Fermi, 5, Firenze, Italy}
\altaffiltext{6}{Dublin Institute for Advanced Studies (DIAS), 31 Fitzwilliam Place, Dublin, Ireland}
\altaffiltext{7}{Departamento de Astronomia, Universidad de Chile, Casilla 36-D, Santiago, Chile}
\altaffiltext{8}{Laboratoire AIM, CEA/IRFU - CNRS/INSU - Université Paris Diderot, CEA-Saclay, 91191 Gif-sur-Yvette Cedex, France}
\altaffiltext{9}{Universite de Bordeaux, OASU, Bordeaux, France}
           

\begin{abstract}
  The ATLASGAL 870\,$\mu$m continuum survey conducted with the APEX
  telescope is the first survey covering the whole inner Galactic
  plane ($60^o>l>-60^o$ \& $b<\pm1.5^o$) in submm continuum emission
  tracing the cold dust of dense and young star-forming regions. Here,
  we present the overall distribution of sources within our Galactic
  disk. The submm continuum emission is confined to a narrow range
  around the galactic plane, but shifted on average by $\sim$0.07\,deg
  below the plane. Source number counts show strong enhancements
  toward the Galactic center, the spiral arms and toward prominent
  star-forming regions.  Comparing the distribution of ATLASGAL dust
  continuum emission to that of young intermediate- to high-mass young
  stellar objects (YSOs) derived from Spitzer data, we find
  similarities as well as differences. In particular, the distribution
  of submm dust continuum emission is significantly more confined to
  the plane than the YSO distribution (FWHM of 0.7 and 1.1\,deg,
  corresponding to mean physical scale heights of approximately 46 and
  80\,pc, respectively). While this difference may partly be caused by
  the large extinction from the dense submm cores, gradual dispersal
  of stellar distributions after their birth could also contribute to
  this effect. Compared to other tracers of Galactic structure, the
  ATLASGAL data are strongly confined to a narrow latitude strip
  around the Galactic plane.
\end{abstract}



\section{Introduction}
\label{intro}

Since the location of our solar system is within the Galactic disk,
studying the Galactic structure of our Milky Way is always a
challenging problem. Therefore, we cannot derive such comprehensive
and intuitive pictures of our disk as extragalactic studies are able
to do for other spiral galaxies (e.g.,
\citealt{kennicutt2003,nieten2006,walter2008}).  Nevertheless, based
on a diverse set of studies over all wavelengths, in the last few
decades we have derived a reasonably comprehensive picture of our
Galactic spiral structure (for recent work, see, e.g.,
\citealt{benjamin2008,reid2009}). The Galactic plane has been observed
in the optical/near-/mid-infrared bands (e.g.,
\citealt{dobashi2005,skrutskie2006,churchwell2009,carey2009}) as well
as at longer wavelengths, e.g., in CO or cm continuum emission (e.g.,
\citealt{dame2001,stil2006}). However, until the arrival of the two
(sub)mm Galactic plane surveys ATLASGAL (The APEX Telescope Large Area
Survey of the GALaxy at 870\,$\mu$m) and BGPS (Bolocam Galactic Plane
Survey) \citep{schuller2009,aguirre2011}, no survey at (sub)mm
wavelengths existed that trace the cold dust emission stemming from
dense and young star-forming regions at adequate spatial resolution
(the COBE (Cosmic Background Explorer) and WMAP (Wilkinson Microwave
Anisotropy Probe) data have too coarse resolution to isolate
individual star-forming regions). Here, we employ the 870\,$\mu$m
submm continuum survey ATLASGAL to study the general distribution of
the dense dust and gas within our Galactic plane.

\section{Observations and source extraction} 
\label{obs}

The 870\,$\mu$m data are taken from the APEX Telescope Large Area
Survey of the Galaxy (ATLASGAL, \citealt{schuller2009}).  The
$1\sigma$ rms of the data is $\sim$50\,mJy\,beam$^{-1}$ and the FWHM
$\sim 19.2''$. Using the clumpfind source identification algorithm by
\citet{williams1994} with a $6\sigma$ threshold of
300\,mJy\,beam$^{-1}$, we identified 16336 clumps within the Galactic
plane for longitudes $60^o>l>-60^o$ and latitudes $b<\pm1.25^o$. In
the context of this paper, we are not aiming for exact fluxes, column
densities or masses, but we just want to evaluate source number counts
within the Galactic plane. Therefore, the specific clump
identification algorithm or the used thresholds are not of great
importance. To test this, we also derived corresponding source
catalogs using $4\sigma$ or $8\sigma$ thresholds. While the absolute
number of sources obviously varies significantly with changed
thresholds, the structural results presented below are not
significantly affected by that. As an additional test, instead of
deriving clumps, we just extracted the total submm fluxes above the
$6\sigma$ threshold in the given latitude and longitude bins.  Again
the structural distributions in longitude and latitude are very
similar. Since other Galactic plane surveys usually also work on
source counts (e.g., GLIMPSE (Galactic Legacy Infrared Midplane
Extraordinaire) or MSX (Midcourse Space Experiment);
\citealt{churchwell2009,robitaille2008,egan2003}), for the remainder
of the paper we adopt the $6\sigma$ source catalog. The clump masses
range between 100 and a few 1000\,M$_{\odot}$, and these clumps form
clusters with certain star formation efficiencies.  Therefore, the
ATLASGAL data largely trace gas/dust clumps capable of forming
intermediate- to high-mass stars at distances between several 100 out
to more than 10000\,pc (\citealt{schuller2009}; Tackenberg et
al.~subm.).

Tackenberg et al.~(subm.) analyzed the ATLASGAL data in the longitude
range between 10 and 20\,deg in depth via correlating them with the
GLIMPSE and MIPSGAL (MIPS Galactic Plane Survey) near- to mid-infrared
surveys of the Galactic plane \citep{churchwell2009,carey2009}, and at
long wavelength with the NH$_3$ spectral line data from Wienen et
al.~(subm.). Out of 210 starless clump candidates, Tackenberg et
al~could extract NH$_3$ spectral information~-- and by that kinematic
distances~-- for 150 sources.  To resolve the kinematic distance
ambiguity, these targets were compared to the GLIMPSE and MIPSGAL
images. Clumps associated with GLIMPSE/MIPSGAL shadows were assigned
the near distance, and the other clumps were assigned as far. This
way, 115 clumps are likely on the near side of the Galaxy and 35 on
the far side.  Tackenberg et al.~find that the mean distances of
starless clumps in the longitude range between 10 and 20\,deg on the
near and far side of the Galaxy are 3.1 and 13.8\,kpc, respectively.
One should keep in mind that the rotation curve of the inner Galaxy is
far from circular (e.g., \citealt{reid2009}), making the absolute
determination of kinematic distances a difficult task.

For comparison with somewhat more evolved evolutionary stages, namely
young stellar objects (YSOs), we resort to the Spitzer red source
catalog (with a color criterion of $[4.5]-[8.0]\geq$1, for more
details see \citealt{robitaille2008}). While the intrinsically red
sources are contaminated by approximately 30\% AGB stars,
\citet{robitaille2008} extracted statistically these AGB stars and
provided a YSO catalog with reduced AGB contamination. This YSO
catalog contains 11649 sources again over the Galactic longitude range
$60^o>l>-60^o$. Because deriving distances and masses for individual
YSOs is difficult, \citet{robitaille2010} conducted a population
synthesis analysis of the sample. They find that their detected
sources consist mainly of intermediate- to high-mass stars (between 3
and 25\,M$_{\odot}$) at distances of several kpc. Again dividing their
sources in near and far sources with respect to the Galactic center
and Galactic bar, they as well find mean distances in the longitude
range between 10 and 20\,deg of 4.9 and 13.1\,kpc, respectively.

\begin{figure}[htb] 
\includegraphics[width=0.48\textwidth]{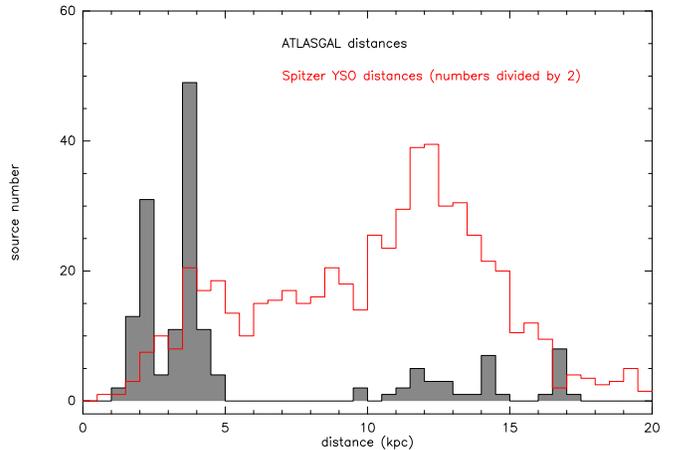}
\caption{Histogram of distances derived for the ATLASGAL clumps
  (black, only the starless clumps) and the GLIMPSE YSOs (red) in the
  Galactic longitude range from 10 to 20\,deg, respectively
  (Tackenberg et al.~subm., \citealt{robitaille2010}).}
\label{distances}
\end{figure}

Figure \ref{distances} presents a histogram of the kinematic distances
in the Galactic longitude range between 10 and 20\,deg derived for the
ATLASGAL sample (Tackenberg et al.~subm.)  and population synthesis
distances for the GLIMPSE sources \citep{robitaille2010},
respectively. Clearly, both distributions show a near and far distance
peak (see also Fig.~9 in \citealt{dunham2011}). For the ATLASGAL
sources, the far peak has less sources because, at the given spatial
resolution of $19.2''$, close clumps that would be spatially resolved
at the near distance merge and appear as one clump at the far side of
the Galaxy (Tackenberg et al.~subm.). To be more precise, Tackenberg
et al.~(subm.) simulated the distance smoothing effect for the
ATLASGAL data, and they found that an artificial sample of 328 clumps
at 3\,kpc distance would appear as only 20 clumps when put to 15\,kpc
distance (see also the corresponding discussion in
\citealt{dunham2011}). This effect is far less severe for the much
better resolved GLIMPSE sources.  Additionally, GLIMPSE sources easier
saturate at the near side of the Galaxy. Combining these effects with
the larger observed volume at the far side of the Galaxy, more sources
are found at the far side for that sample.  Although the near peaks of
the ATLASGAL and GLIMPSE samples are shifted with respect to each
other a little bit, taking into account the inherent uncertainties of
kinematic distances for the ATLASGAL sample and population synthesis
for the YSO sample ($\sim$1\,kpc each), the two source types cover
comparable distance ranges.  Another difference between the two
samples is that ATLASGAL barely detects any sources between 5 and
10\,kpc whereas the GLIMPSE distribution shows sources there. The
latter is due to the model used for the population synthesis with an
axisymmetric ring around the center of the Galaxy (see Fig.~2 in
\citealt{robitaille2010}).  The non-detection of ATLASGAL sources in
that distance regime is likely attributed to the large extend of the
Galactic bar (see also Fig.~\ref{sketch}) which is not taken into
account in Galactic rotation models to derive the kinematic distances
(e.g., \citealt{reid2009}).

While the ATLASGAL clumps are in absolute terms more massive than the
GLIMPSE YSOs, considering typical star formation rates and an initial
mass function for each cluster-forming region, the ATLASGAL clumps
form intermediate- to high-mass stars in the mass range of the GLIMPSE
YSOs.

In combination, the two samples are well suited for comparison of the
different evolutionary populations. In addition to this, since the
Galactic longitude range between 10 and 20 represents a fraction of
the Galactic plane that covers near and far spiral arms, we consider
it to first order as representative to extrapolate the distance
similarities for both samples also for the rest of the surveys.
Additional similarities and differences will be discussed in section
\ref{lat}.

\section{Results}
\label{results}

\begin{figure*}
\includegraphics[width=0.98\textwidth]{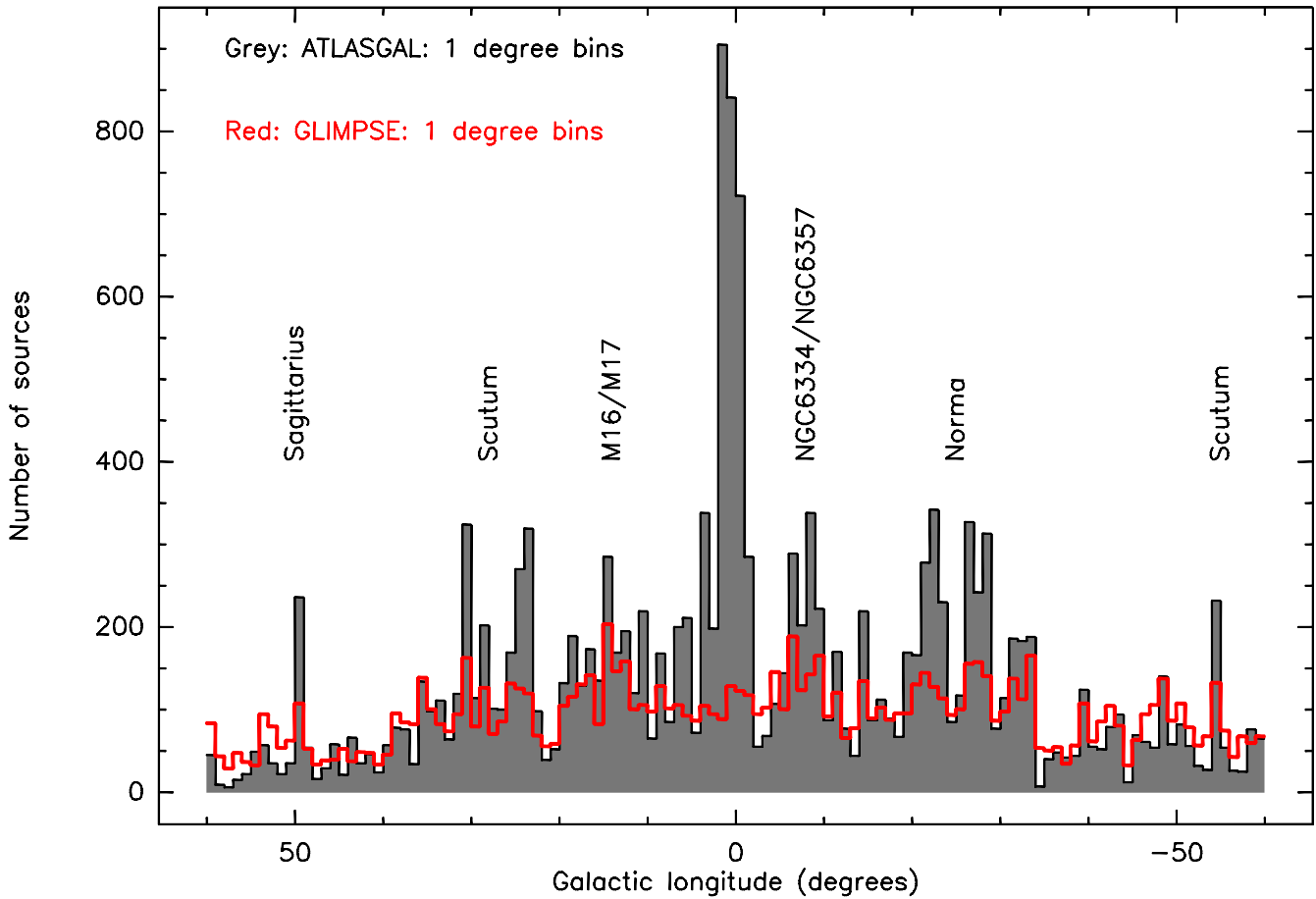}
\caption{Histogram of source number counts with Galactic longitude.
  The grey-scale shows the ATLASGAL submm continuum sources, and the
  red histogram presents the YSOs derived from the Spitzer data
  \citep{robitaille2008}. The data are binned in longitude in 1\,deg
  bins. For ATLASGAL we use the data for latitudes between $\pm
  1.25$\,deg whereas the GLIMPSE YSO data are restricted to latitudes
  between $\pm 1.0$\,deg.}
\label{histo_lon}
\end{figure*}

\begin{figure}
\includegraphics[width=0.48\textwidth]{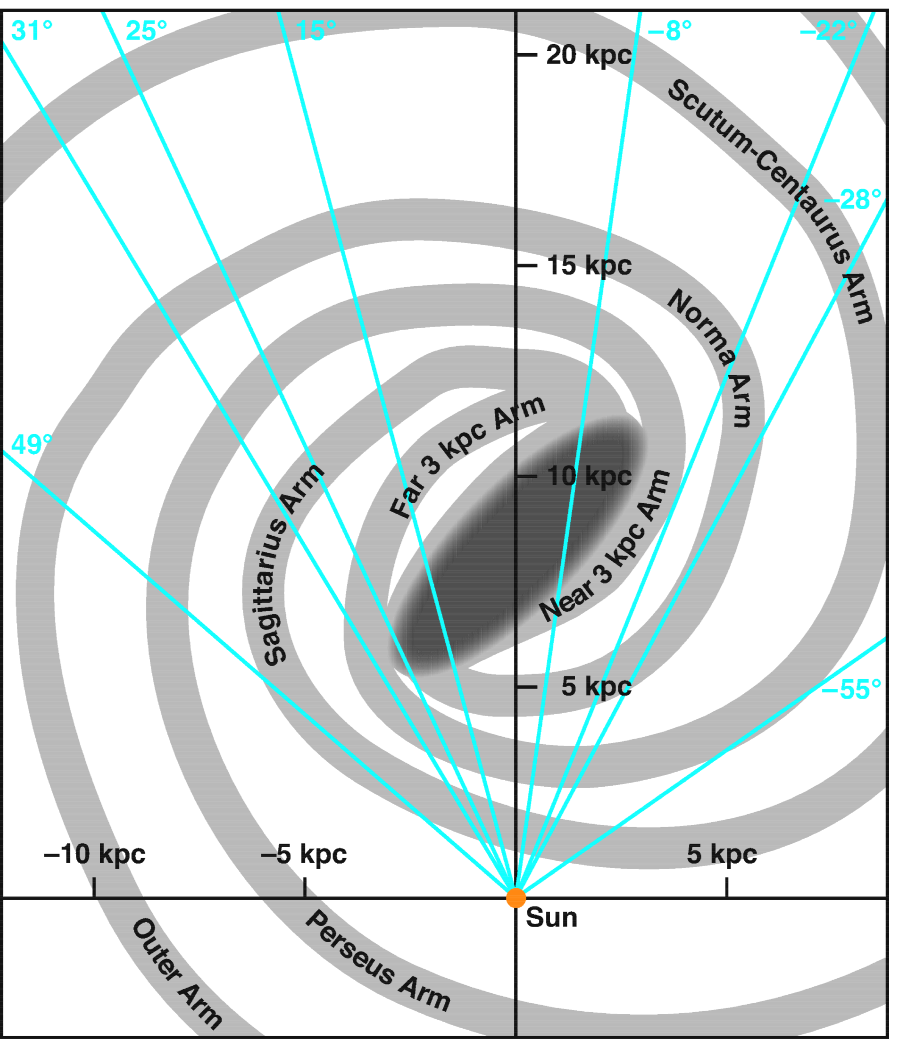}
\caption{Sketch of the Galactic plane with several prominent lines of
  sight marked. Artist impression (by MPIA graphics department) of
  face-on view of the Milky Way following the Galactic structure
  discussed in \citet{reid2009}.}
\label{sketch}
\end{figure}

Figure \ref{histo_lon} presents the longitude distribution of the
submm continuum and YSO sources within our Galactic plane (see
\citet{schuller2009} for a first version of such a plot based on a far
smaller initial dataset). While the YSO distribution based on the
Spitzer data is relatively flat, the submm continuum emission shows a
series of distinctive peaks. The most prominent one is toward the
Galactic center where the source count increases approximately by a
factor of 4. In addition to this, there are a few more clear submm
source count peaks at positive and negative longitudes. They can
mainly be attributed to tangential points of spiral arms (for a
discussion of older COBE 240\,$\mu$m data, see \citealt{drimmel2000})
as well as to a few prominent star formation complexes in the nearby
Sagittarius arm. The most important of these are marked in Figure
\ref{histo_lon}. For comparison, a sketch of our Galaxy as it would be
viewed face-on is shown in Fig.~\ref{sketch} where several lines of
sight are marked corresponding to increased number counts in
Fig.~\ref{histo_lon}. For longitudes $> -10$\,deg, a similar
distribution was found in the BGPS survey \citep{rosolowsky2010}. The
implications suggested by this result will be discussed further in
section \ref{general}.

\begin{figure}
\includegraphics[width=0.48\textwidth]{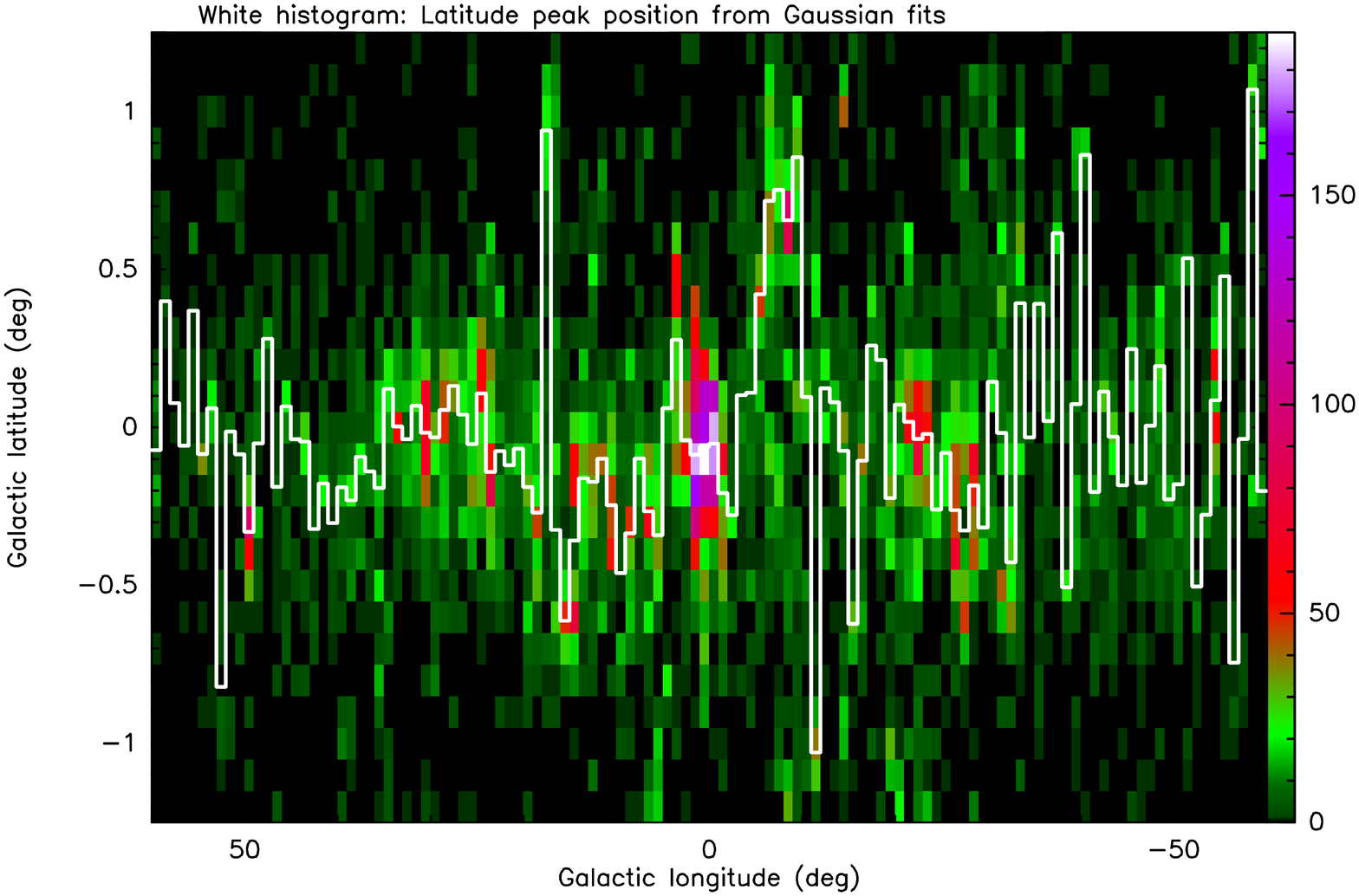}\\
\includegraphics[width=0.48\textwidth]{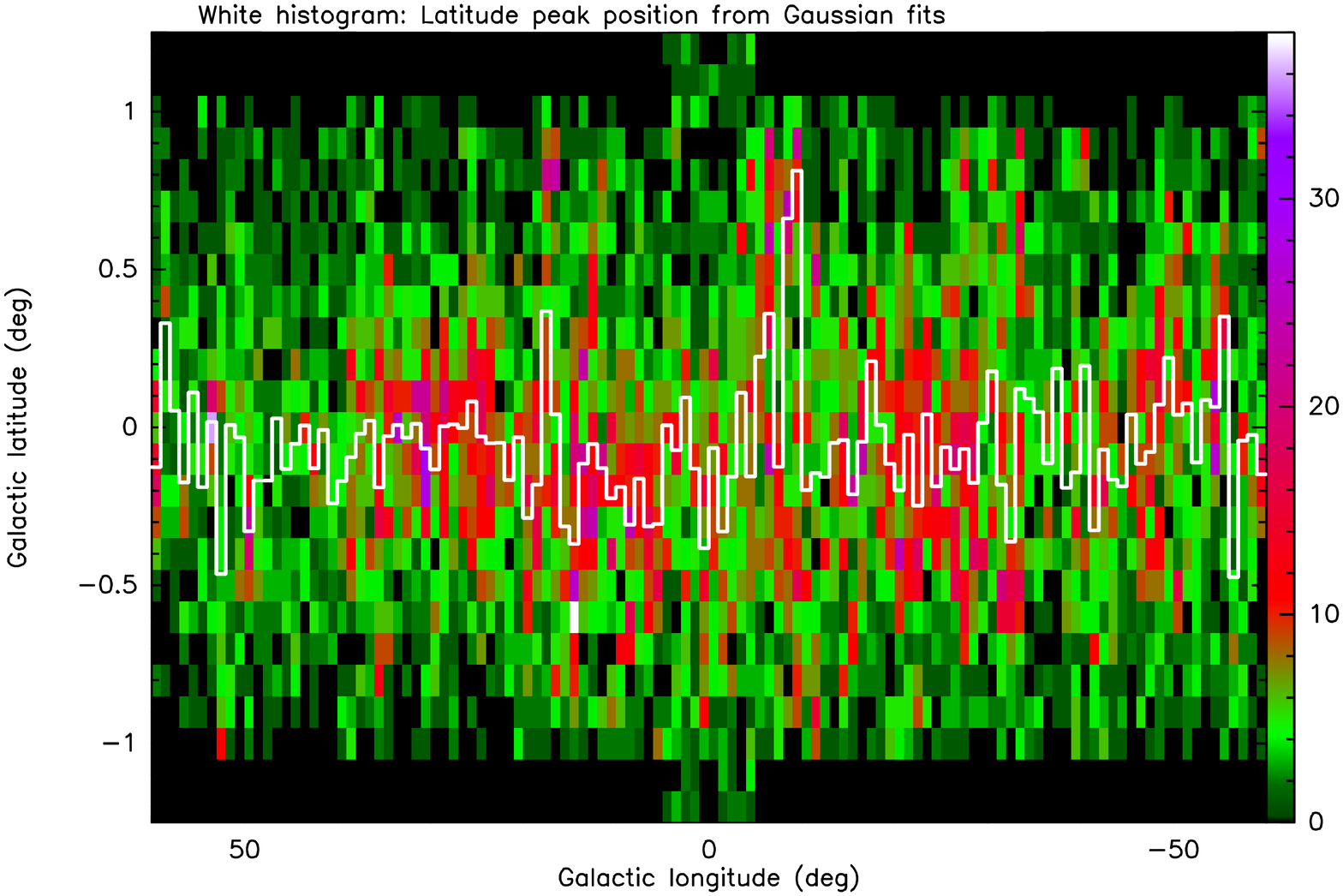}
\caption{The color-scale shows the two-dimensional source count
  distribution for ATLASGAL submm continuum (top panel) and GLIMPSE
  YSO (bottom panel) sources. The bin sizes in Galactic longitude and
  latitude are 1 and 0.1\,deg, respectively. The white lines mark the
  peak positions of Gaussian fits to the latitude distributions at
  each given longitude.}
\label{lon_lat}
\end{figure}

Another way to represent the 870\,$\mu$m source distribution is in a
2-dimensional binning in Galactic longitude and latitude
(Fig.~\ref{lon_lat} top panel). In addition to an increase in source
counts toward specific Galactic longitudes, we also identify a tight
confinement to the Galactic mid-plane with only a narrow spread north
and south of the mid-plane. To derive the approximate latitude for which
the submm emission peaks in each longitude bin, we fitted Gaussians to
their latitude distribution in each longitude bin. The Gaussian fit
peak positions are marked in Fig.~\ref{lon_lat} (top panel). These
fits indicate that the dominant dust and gas distribution is slightly
shifted to negative Galactic latitudes with a mean offset over the
whole plane of $-0.076\pm 0.008$\,deg (the mean values are derived from
Gaussian fits to 10\,deg longitude bins, see below).

\begin{figure}
\includegraphics[width=0.48\textwidth]{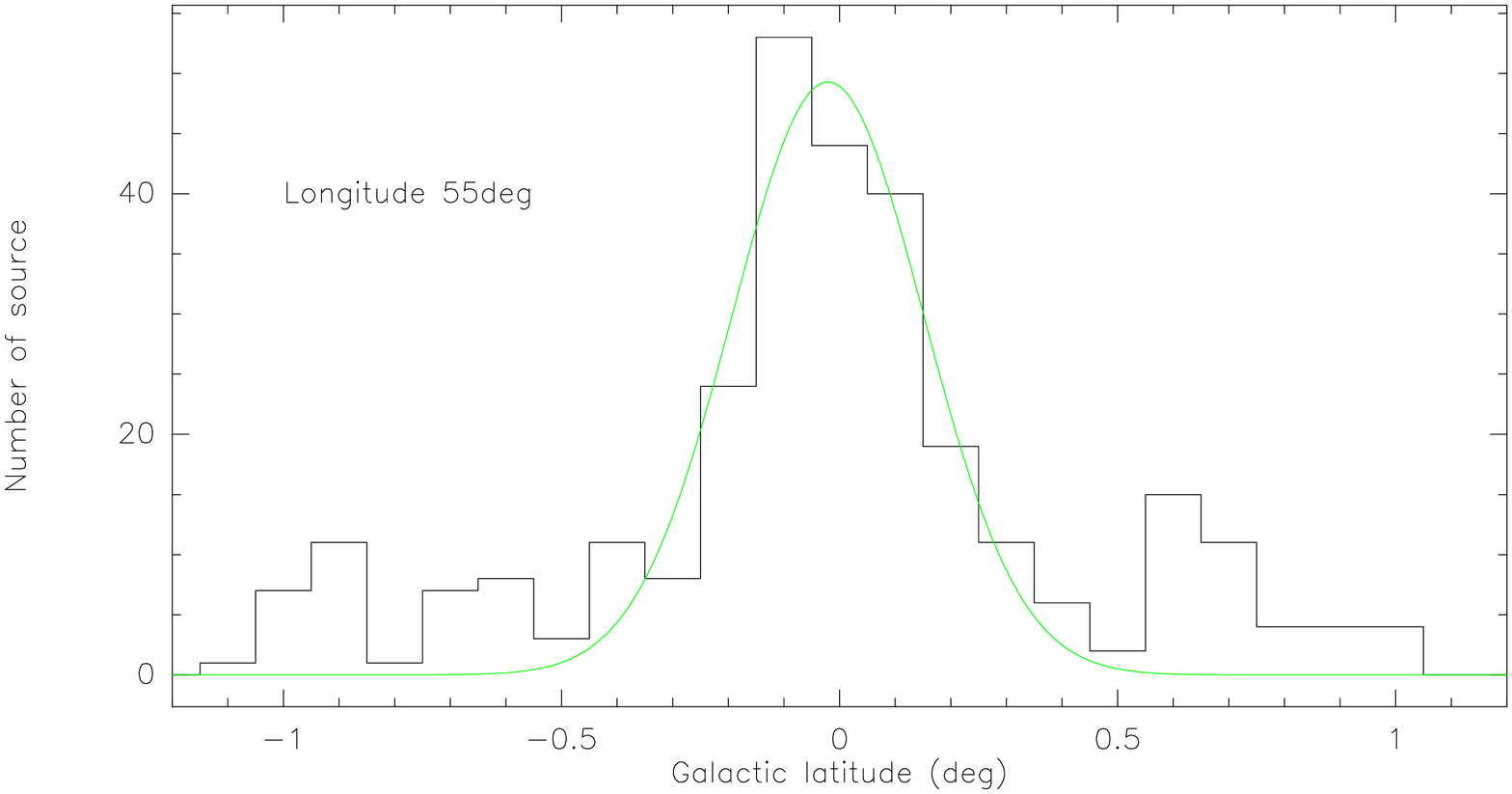}\\
\includegraphics[width=0.48\textwidth]{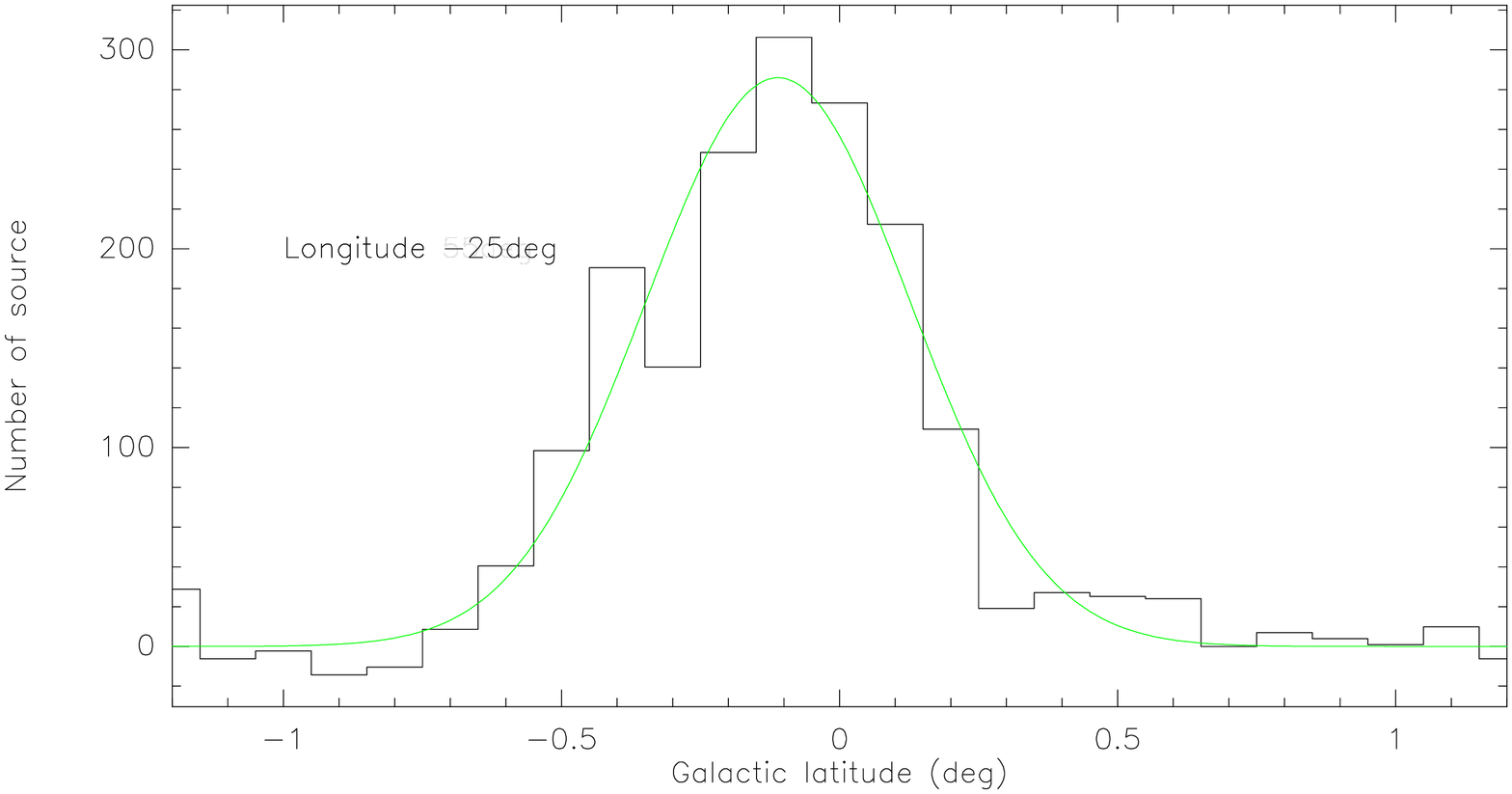}
\caption{Examples of the Gaussian fits to the latitude distributions.
  The centers of the 10\,deg longitude bins are labeled in each
  panel.}
\label{gauss}
\end{figure}

\begin{figure}
\includegraphics[width=0.48\textwidth]{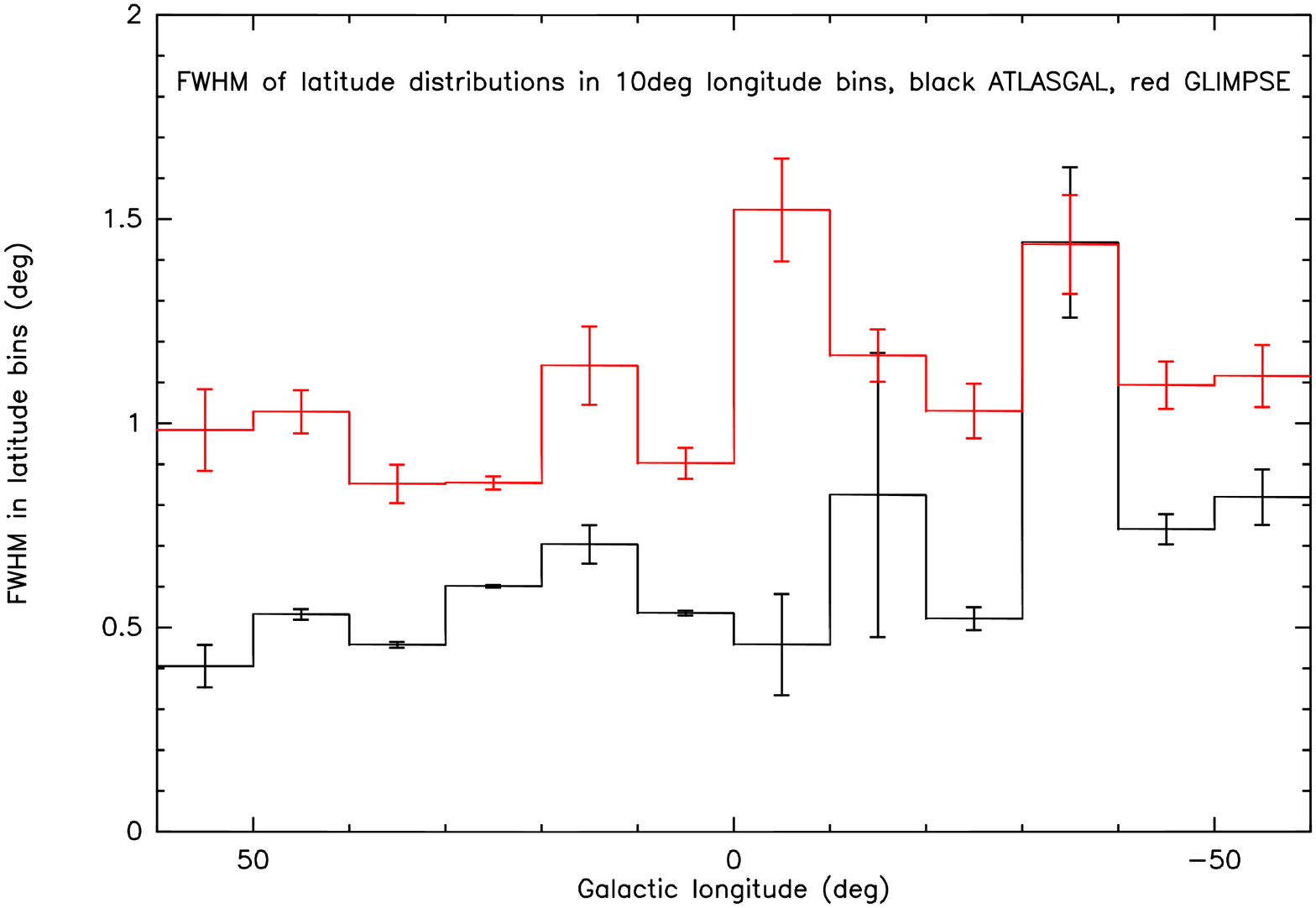}\\
\caption{Resulting FWHM of Gaussian fits (and associated errors) to
  the latitude distribution in 10\,deg Galactic longitude bins. The
  black histogram is from the ATLASGAL data, and the red histogram
  from the GLIMPSE data.}
\label{width}
\end{figure}

We can produce the same plot for the YSO distribution derived from the
Spitzer data (Fig.~\ref{lon_lat} bottom panel). Similarly, the mean
value of the peaks of the YSO distribution is also shifted below the
Galactic mid-plane, again at $-0.072\pm 0.008$\,deg (the mean values
are also derived from Gaussian fits to 10\,deg longitude bins).
Already a visual inspection of the two distributions indicates that
the YSOs appear to cover a broader range in Galactic latitude than the
dust and gas clumps traced by the submm continuum emission.  Fitting
Gaussians to the latitude distributions in each longitude bin for the
submm clumps as well as the YSOs allows us to better quantify this
effect.  Since the latitude distributions are not as smooth on the
scales of individual degrees in Galactic longitude, we average over 10
deg in longitude for smoothing purposes. Figure \ref{gauss} presents
Gaussian example fits at different Galactic longitudes outlining the
applicability of the Gaussian assumption to these distributions.
%  One should note that Gaussian distributions
%  with the same scale covering distances varying by a factor of a few
%  would result in a convolved non-Gaussian distribution. The fact that
%  the observed distributions resemble Gaussians relatively well
%  indicates that in many cases the distributions should be dominated
%  by one distance. Independent of that, 
The corresponding Gaussian full width half maximum (FWHM) for the two
distributions are shown in Figure \ref{width}. One clearly sees that
the YSO distributions is broader over the whole Galactic plane than
the dense gas and dust distribution. Below Galactic longitudes of
-30\,deg, the ATLASGAL distribution shows a tendency of increased
FWHM. However, this effect is confined to only three bins, one with a
particularly large error-bar. Therefore, in the context of this paper
we refrain from further interpretation. The mean values of the FWHM
for the dust continuum and YSO distributions are $\sim$0.67$\pm 0.02$
and $\sim$1.09$\pm 0.02$\,deg, respectively. This corresponds to
characteristic scale heights $H$ (distance where distribution has
dropped to 1/e, $H\approx$0.6$\times$FWHM) of $\sim$0.4 and
$\sim$0.7\,deg for the submm continuum and YSO distributions,
respectively. \citet{robitaille2010} fitted their source distribution
with a mean physical scale height of 80\,pc.  Since we have no
explicit distances for individual ATLASGAL sources over the whole
range of the Galactic plane, deriving directly a physical scale height
from our data alone is hardly feasible.  However, because the distance
and mass regimes of the ATLASGAL sources and the YSOs are similar (see
section \ref{obs}), we can use the linear scale height of the YSOs to
derive a first-order estimate of the linear scale height for the
ATLASGAL clumps (see section \ref{lat} for additional discussion).
With the measured FWHMs of the YSOs and the ATLASGAL sources, and
using the modeled 80\,pc mean physical scale height of the YSOs, we
estimate, proportionally, an approximate physical scale height of the
submm dust continuum emission of 46\,pc. Since the distribution of
near and far sources favors near sources for ATLASGAL and far sources
for GLIMPSE (see Fig.~\ref{distances}), the 46\,pc should be
considered as an upper limit.

Another estimate of the ATLASGAL scale height can be derived from the
sample discussed in section \ref{obs} for the Galactic longitude range
between 10 and 20\,deg (Tackenberg et al.~subm). Using only the near
kinematic distance sample (the far distance sample has too low number
statistics), we find a FWHM and scale height of 23 and 14\,pc,
respectively. This is considerably lower than the estimate of 46\,pc
derived above. The most important reason for this difference is based
on the way near and far distance sources are separated in the study by
Tackenberg et al.~(subm.): While the kinematic distances always give
near and far solutions at this and the other side of the Galaxy, the
near solution was favored for infrared dark clouds if absorption
shadows were detected in the Spitzer mid-infrared data (e.g.,
\citealt{peretto2009}). However, at high Galactic latitudes the
infrared background is significantly reduced and by default it is more
difficult to identify infrared dark clouds. Therefore, from a
statistical point of view, at high latitude, clumps are potentially
shifted to the far distance. This implies that the derived scale
height of 14\,pc from this sample has to be considered as a lower
limit.

\section{Discussion}
\label{general}

\subsection{Longitude distribution}
\label{lon}

To first order, it appears surprising that the cold dust clump
distribution and the YSO distribution do not resemble each other more
closely. While the submm continuum emission is optically thin and
should therefore trace almost all cold dust along each line of sight,
the YSO distribution is more strongly affected by extinction. It may
well be that a considerable fraction of YSOs are not identified
because of too high extinction. Regarding the different properties of
the ATLASGAL clumps and YSOs toward the Galactic center
(Fig.~\ref{histo_lon}), a similar emission increase toward the
Galactic center was recently also found in the dense gas emission of
NH$_3$ (the HOPS survey, e.g., \citealt{walsh2011}, Purcell et
al.~subm., Longmore et al.~in prep.).  The lack of a prominent peak
toward the Galactic center in the GLIMPSE YSO data may be partly an
observational bias but also partly a real physical effect.
Observationally, the extinction toward the Galactic center increases
which increases the GLIMPSE detection threshold.  However, in contrast
to that, also surveys of H{\sc ii} regions, tracing more evolved
high-mass star-forming regions, as well as H$_2$O and CH$_3$OH maser
surveys exhibit no strong emission peaks toward the Galactic center
(e.g.,
\citealt{wilson1970,lockman1979,bronfman2000,anderson2011,walsh2011,green2011}).
Does that imply a relatively low gas-to-star conversion in that
specific part of our Galaxy? For a more detailed discussion, see
Longmore et al.~in prep..

\citet{simon2006} presented a similar plot to our Fig.~\ref{histo_lon}
for the distribution of IRDCs in the Galactic plane. While the general
features for the IRDCs appear similar to the dust continuum
distribution, the spiral arm and star-forming regions are less
pronounced. \citet{anderson2011} performed a similar study of the
H{\sc ii} region distribution of the Galactic plane accessible to the
northern hemisphere, whose results are noticeably different to what we
find. The H{\sc ii} region distribution does not exhibit a peak toward
the Galactic center region, but a clear peak is found at approximately
+30\,deg, corresponding to our peak for the Scutum arm.  Previous
H{\sc ii} region surveys show a similar H{\sc ii} number increase at
negative longitudes around -30\,deg
\citep{wilson1970,lockman1979,bronfman2000}.  

Recently, \citet{green2011} report on the CH$_3$OH maser distribution
in the Galactic plane between longitude $\pm 28$\,deg.  Among other
source count peaks, in particular they report an increased detection
rates at longitudes around $+25$ and $-22$\,deg, very similar to what
we find. While the peaks at $+25$ and $+31$ deg are likely associated
with the end of the long Galactic bar and the beginning of the
Scutum-Centaurus spiral arm, the $-22$\,deg peak should be associated
with a tangent point of the 3\,kpc arm (Fig.~\ref{sketch}). In
summary, the submm continuum emission is an excellent tracer of the
Galactic dense gas structure, even in our Milky Way where our location
within the plane complicates the picture so severely.

\subsection{Latitude distribution}
\label{lat}

The finding that the average peak of the Gaussian latitude
distribution is below the Galactic plane has already been inferred by
other groups, e.g., for (sub)mm continuum emission
\citep{schuller2009,rosolowsky2010}, in the infrared
\citep{churchwell2006,churchwell2007,robitaille2010}, for CO emission
\citep{cohen1977}, clusters \citep{mercer2005}, H{\sc ii} regions
\citep{lockman1979,bronfman2000} or infrared bubbles (Kendrew
priv.~comm., Simpson et al.~in prep.). Even the Galactic center itself
is located at 0.05\,deg below the Galactic plane \citep{reid2004}.
While a common interpretation of that effect is based on a poor
definition of the Galactic plane where neither the sun nor the
Galactic center itself are located directly in the plane at 0\,deg
latitude (e.g., \citealt{humphreys1995,joshi2007,schuller2009}),
\citet{rosolowsky2010} recently suggested that this effect may also
simply be caused by individual star formation complexes and not so
much reflect a global Galactic property. However, they state that the
offset is mainly a feature of the molecular interstellar medium,
whereas different studies of the GLIMPSE survey indicate that the
stellar component shows the same effect (e.g.,
\citealt{mercer2005,churchwell2006}, this study).  Although we cannot
conclusively differentiate these scenarios, the data here are
indicative of a real global offset of the Galactic midplane from its
conventional position where the axis between the sun and the Galactic
center are located at $b=0$\,deg.

Are the different Galactic latitude distributions of the submm clumps
and the YSOs a real physical effect or could they be caused by
observational biases?  As outlined in section \ref{obs}, the mass
distributions and the distances on the near and far side of the Galaxy
of the two samples are similar. One may now ask whether the number of
near sources were larger for the YSOs than for the submm clumps.
However, there are several effects that counteract this: At the
relatively coarse spatial resolution of ATLASGAL ($19.2''$), clumps
that would be separate entities on the near side of the Galaxy merge
into single objects on the far side, and the total number of sources
on the far side is lower than that on the near side (Tackenberg et
al.~subm.).  In contrast to that, at the higher spatial resolution of
Spitzer ($2''$), most sources remain point sources independent of the
distance, and therefore less suffer from the ``merging-problem''.
Furthermore, the observed volume at the far side of our Galaxy is
larger than that on the near side, and GLIMPSE sources on the near
side more easily saturate.  These combined effects even cause a YSO
number increase on the far side compared to the near side of our
Galaxy (Fig.~\ref{distances}, \citealt{robitaille2010}).  Therefore,
the mass and distance distributions of the submm clumps and YSOs are
unlikely to be the cause for the difference in the latitude
distribution.  Furthermore, \citet{robitaille2008} statistically
excluded the AGB star population from their catalogue, which implies
that contamination by post-main-sequence sources is not responsible
for the difference in latitude distribution as well.  Similar to the
effect seen for the longitude distribution discussed in the previous
section (\S \ref{lon}), extinction may cause part of the broader
latitude distribution because toward the highest extinction regions,
that are traced by the submm continuum emission, infrared emission is
hardly detectable and hence the most deeply embedded YSO population is
likely to be missed by the Spitzer data or the selection criteria in
\citet{robitaille2008}.  Herschel longer-wavelength data may identify
such embedded population better (e.g.,
\citealt{hennemann2010,henning2010,beuther2010b}).

\citet{walsh2011} recently reported a Galactic average scale height
for H$_2$O masers of approximately 0.4\,deg, earlier finding a scale
height for the CH$_3$OH class {\sc II} masers in a similar range
\citep{walsh1997}. Assuming a comparable distance distribution for the
maser as well as the submm continuum sources, the mean physical
scale-height of the masers should also be $\approx$46\,pc.  For
ultracompact H{\sc ii} regions, \citet{wc89} found a scale height of
0.6\,deg, intermediate between our values for the submm clumps and the
YSOs. Later, \citet{becker1994} reported a smaller mean physical
scale-height for ultracompact H{\sc ii} regions of $\sim$30\,pc (about
40\% of the \citealt{wc89} value), claiming that the \citet{wc89}
sample is biased by its large fraction of B-stars. Similar mean
physical scale heights for high-mass star-forming regions of
$\sim$44\,pc and $\sim$29\,pc were reported by \citet{bronfman2000}
and \citet{urquhart2011}.  Hence masers as typical tracers of
star-forming regions (a fraction of the H$_2$O masers may also stem
from AGB stars), dust continuum emission as a tracer of the dense gas,
and young high-mass stars exhibit similar scale height distributions
in the Milky Way. In comparison, \citet{bronfman1988} find an
approximate scale height of 70\,pc for CO (rescaled to a
galactocentric solar distance of 8.5\,kpc), and \citet{dame1994}
derive a values of $\sim$120\,pc for the thick CO disk (which is an average of
their 3 cited values).
%Comment Leo Bronfman: The thickness quoted is of 70 pc HWHM. That
%makes a FWHM of 140 pc, and a physical height scale of 0.6*140 = 84
%pc.  But the galactocentric solar distance used at that time was Ro =
%10 kpc. The number has to be scaled to the value Ro = 8.5 kpc, in
%present use.  Therefore the mean physical scale height for CO to be
%quoted is of 0.85* 84 = 71 pc (~ 70 pc)
The reported cold HI scale height is around 150\,pc
\citep{kalberla1998,kalberla2003,dedes2005}. Therefore, while tracers
of the youngest evolutionary stages of star formation (submm continuum
emission and masers) are all found closest to the Galactic plane, more
evolved evolutionary stages like YSOs as well as the less dense atomic
and molecular gas appear to be located on average slightly further from
the plane.

Using the estimated mean physical scale heights for the dust continuum
and YSO distributions of 46 and 80\,pc, respectively, we can calculate
approximate velocities required to move the 30\,pc difference in the
given YSO lifetimes of 1--2\,Myrs. This estimate results in required
YSO velocities between 15 and 30\,km\,s$^{-1}$. While velocities of
that order are found (e.g., PV Cephei, \citealt{goodman2004}), they
are apparently not the rule. Therefore, we propose that the most
likely explanation for the spread in scale height for different
populations appears to be a combination of extinction effects and
dissolving young clusters from their natal birth sites.

\section{Conclusions}

We present a study of the Galactic distribution of submm dust
continuum emission from the northern and southern hemisphere. The
submm continuum emission, which traces the dense gas emission from
star-forming regions, excellently traces the structure within our
Galaxy. Spiral arms and prominent structures of star formation are
easily distinguished by significantly increased source counts toward
these Galactic longitudes. The Galactic latitude distribution is
skewed slightly below the Galactic plane, and the mean physical scale
height is estimated to 46\,pc. This scale height corresponds well to
other star-formation tracers like CH$_3$OH and H$_2$O maser emission,
and it is more confined to the Galactic plane than most of the other
populations in our Galaxy. We compare the submm continuum emission
with several other tracers of Galactic structure, in particular with
the YSO population identified by Spitzer observations. The YSO
population has a significantly larger scale height, and we propose
that this may be attributed to a combined effect of extinction and
dissolving of protostellar clusters after their birth.

\begin{acknowledgements} 
  We thank the referee for a careful review improving the paper
  significantly.  H.B. likes to thanks Sarah Kendrew for discussion
  about Galactic distributions. LB acknowledges support from CONICYT
  through projects FONDAP No. 15010003 and BASAL PFB-06.
\end{acknowledgements}

%\bibliography{/home/beuther/tex/bibliography}   
%\bibliography{/Users/henrikbeuther/paper/bibliography}
%\bibliographystyle{aa}    % this does the style, aa.bst necessary
%\input{ms.bbl}

\end{document}